\documentclass[a4paper,11pt]{article}
\pdfoutput=1
\usepackage{jcappub}
\usepackage[T1]{fontenc}

\usepackage{tikz}
\usepackage{tikz-3dplot}
\usepackage{enumitem}
\usepackage{mathtools}
\usepackage{array}
\usepackage{colortbl}

\title{Impact of half-wave plate systematics on the measurement of cosmic birefringence from CMB polarization}

\author[a]{Marta Monelli,}
\author[a,b]{\!Eiichiro Komatsu,}
\author[c]{\!Alexandre E.\ Adler,}
\author[d,e,f]{\!Matteo Billi,}
\author[a,g]{\!Paolo Campeti,}
\author[c]{\!Nadia Dachlythra,}
\author[h, i]{\!Adriaan J.\ Duivenvoorden,}
\author[c, j]{\!Jon E.\ Gudmundsson,}
\author[a]{and Martin Reinecke.}

\affiliation[a]{
Max Planck Institute for Astrophysics, Karl-Schwarzschild-Str.\ 1, 85748 Garching, Germany
}
\affiliation[b]{
Kavli Institute for the Physics and Mathematics of the Universe (Kavli IPMU, WPI), UTIAS, The University of Tokyo, Chiba, 277-8583, Japan
}
\affiliation[c]{
The Oskar Klein Centre, Department of Physics, Stockholm University, SE-106 91 Stockholm, Sweden
}
\affiliation[d]{
Instituto de Física de Cantabria (IFCA), CSIC-UC, Avenida de Los Castros s/n, 39005 Santander, Spain
}
\affiliation[e]{
Dipartimento di Fisica e Astronomia, Alma Mater Studiorum Università di Bologna, Via Gobetti 93/2, I-40129 Bologna, Italy
}
\affiliation[f]{
Istituto Nazionale di Astrofisica - Osservatorio di Astrofisica e Scienza dello Spazio di Bologna, via Gobetti 101, I-40129 Bologna, Italy
}
\affiliation[g]{
Excellence Cluster ORIGINS, Boltzmannstr.\ 2, 85748 Garching, Germany
}
\affiliation[h]{
Center for Computational Astrophysics, Flatiron Institute, 162 5th Avenue, New York, NY 10010, USA}
\affiliation[i]{
Joseph Henry Laboratories of Physics, Jadwin Hall, Princeton University, Princeton, NJ 08544, USA
}
\affiliation[j]{
Science Institute, University of Iceland, Dunhaga 3, IS-107 Reykjavik, Iceland
}

\emailAdd{monelli@mpa-garching.mpg.de}

\abstract{
Polarization of the cosmic microwave background (CMB) can probe new parity-violating physics such as cosmic birefringence (CB), which requires exquisite control over instrumental systematics. 
The non-idealities of the half-wave plate (HWP) represent a source of systematics when used as a polarization modulator.
We study their impact on the CMB angular power spectra, which is partially degenerate with CB and miscalibration of the polarization angle. 
We use full-sky beam convolution simulations including HWP to generate mock noiseless time-ordered data, process them through a bin averaging map-maker, and calculate the power spectra including $TB$ and $EB$ correlations.
We also derive analytical formulae which accurately model the observed spectra. 
For our choice of HWP parameters, the HWP-induced angle amounts to a few degrees, which could be misinterpreted as CB. Accurate knowledge of the HWP is required to mitigate this. Our simulation and analytical formulae will be useful for deriving requirements for the accuracy of HWP calibration.
}

\notoc

\begin{document}

\maketitle
\flushbottom

\section{Introduction}
Temperature anisotropies in the cosmic microwave background (CMB) are an invaluable source of cosmological information \cite{BOOMERANG:2005,Komatsu:2014ioa,Planck:2018cosmpar}. Polarization anisotropies also contain a great wealth of complementary information \cite{WMAPmaps,Planck:2018nkj,POLARBEAR:2019kzz,Polarbear:2020lii,ACT:2020gnv,SPT:2019nip,SPT-3G:2021eoc,BICEP:2021xfz,SPIDER:2021ncy}, which has yet to be fully explored. A promising opportunity driving the development of a major experimental effort, involving both ground-based observatories (Simons Observatory \cite{Ade_2019}, South Pole Observatory \cite{Moncelsi:2020ppj} and CMB Stage-4 \cite{Abazajian:2019eic}) and space missions (LiteBIRD \cite{LiteBIRD:2022cnt} and PICO \cite{NASAPICO:2019thw}), is to probe cosmic inflation \cite{PhysRevD.23.347,Sato:inflation,Linde:1981mu}.
Inflationary models predict the existence of a stochastic background of gravitational waves \cite{Grishchuk:1974ny,Starobinsky:1979ty} which would leave a distinctive $B$-mode signature on the CMB polarization \cite{Zaldarriaga:1996xe,Kamionkowski:1996ks,Seljak:1996gy,Kamionkowski:1996zd}.

The CMB polarization can also probe new parity-violating physics \cite{Komatsu:2022nvu}. For example, in the presence of a time-dependent parity-violating pseudoscalar field, the linear polarization plane of CMB photons would rotate while they travel towards us \cite{Carroll:1989vb,PhysRevD.43.3789,Harari:1992ea}. Because of its similarity with photon propagation through a birefringent material, this phenomenon is referred to as cosmic birefringence (CB). The so-called CB angle, $\beta$, denotes the overall rotation angle from last scattering to present times.  Although the effect of $\beta$ on the observed CMB angular power spectra is degenerate with an instrumental miscalibration of the polarization angle \cite{QUaD:2008ado,WMAP:2010qai,Keating:2012ge,LiteBIRD:2021hlz}, the methodology proposed in \cite{Minami:2019ruj,Minami:2020xfg,Minami:2020fin}, which relies on the polarized Galactic foreground emission to determine miscalibration angles, allowed to infer $\beta = 0.35 \pm0.14^\circ$ at $68\%$ C.L. \cite{Minami:2020odp} from nearly full-sky \emph{Planck} polarization data \cite{Planck:2018lkk}. Subsequent works \cite{Diego-Palazuelos:2022dsq,Eskilt:2022wav,Eskilt:2022cff} reported more precise measurements for $\beta$. The statistical significance of $\beta$ is expected to improve with the next generation of CMB experiments, given the high precision at which they aim to calibrate the absolute position angle of linear polarization. This will make it unnecessary to rely on the Galactic foreground to calibrate angles and measure $\beta$ \cite{Komatsu:2022nvu}, hence avoiding the potential complications highlighted in \cite{Clark:2021kze}.

The unprecedented sensitivity goals of future surveys, aiming to detect faint primordial $B$ modes, can only be achieved if systematics are kept under control. To this end, a promising strategy is to employ a rotating half-wave plate (HWP) as a polarization modulator. As shown by the previous analyses \cite{Johnson_2007,2010SPIE.7741E..1CR,ABS:2013dqh,Rahlin:2014rja,Misawa:2014hka,Hill:2016jhd,Takakura:2017ddx,2016SPIE.9914E..0JG}, a rotating HWP can both mitigate the $1/f$ noise component \cite{Johnson_2007} and reduce a potential temperature-to-polarization ($I\to P$) leakage due to the pair differencing of orthogonal detectors \cite{Bryan:2015qwa,2016RScI...87i4503E}. Because of these advantages, HWPs are used in the design of some next-generation experiments, including LiteBIRD \cite{LiteBIRD:2022cnt}. However, non-idealities in realistic HWPs induce additional systematics which should be well understood in order for future experiments to meet their sensitivity requirements. This necessity motivated a number of recent works, from descriptions of HWP non-idealities \cite{2010SPIE.7741E..2BB,ABS:2013dqh,pisano2014development,2017arXiv170602464A} and their impact on measured angular power spectra \cite{Giardiello:2021uxq} to mitigation strategies \cite{2012ApJ...747...97B,Matsumura:2014dda,Bao:2015eaa,Verges:2020xug}.

In this paper we study how HWP non-idealities can affect the estimated CMB angular power spectra if overlooked in the map-making step. We employ a modified version of the publicly available beam convolution code \texttt{beamconv}\footnote{\url{https://github.com/AdriJD/beamconv}} \cite{Duivenvoorden:2018zdp,Duivenvoorden:2020xzm} and simulate two sets of noiseless time-ordered data (TOD). The two simulations make different assumptions on the HWP behavior. In the first case the HWP is assumed to be ideal, while non-idealities are included in the second case. We then process the two TOD sets with a map-maker assuming the ideal HWP and compare the output power spectra.
We also derive a set of analytic expressions for the estimated angular power spectra as functions of the input spectra and the elements of the HWP Mueller matrix. These formulae accurately model the output power spectra.
Finally, we show that neglecting the non-idealities in the map-maker affects the observed spectra in a way that is partially degenerate with the CB and instrumental miscalibration of the polarization angle. This effect is evident in the simulations and the analytical formulae. 

The rest of this paper is organized as follows.
In section \ref{sec:math} we present a simple data model for the signal measured by a single detector; generalize it to a larger focal plane and a longer observation time; and introduce the bin averaging map-making method employed in the paper to convert the TOD to maps. In section \ref{sec:simulation} we discuss the instrument specifics we have implemented in the simulation and show the output angular power spectra. The interpretation of the result is the topic of section \ref{sec:analytical}, where we derive some analytical formulae modeling it with good precision. In section \ref{sec:impact} we show how the effect of the HWP non-idealities is partially degenerate with an instrumental miscalibration of the polarization angle, and can therefore be misinterpreted as CB. We quantify the HWP-induced miscalibration angle, which amounts to a few degrees for our choice of the HWP parameters. Conclusions and outlook are presented in section \ref{sec:conclusions}.

\section{Mathematical framework: data model and map-maker}\label{sec:math}
\paragraph{Data model for a single detector}
Polarized radiation can be described by the Stokes $I$, $Q$, $U$ and $V$ parameters or, more compactly, by a Stokes vector, $\mathbf{S}\equiv(I,Q,U,V)$. 
In this paper we use the ``CMB convention'' for the sign of Stokes $U$ \cite{diSeregoAlighieri:2016lbr} and define the Stokes parameters in right-handed coordinates with the $z$ axis taken in the direction of the observer's line of sight (telescope boresight).
The Stokes vector is transformed as $\mathbf{S}\to \mathbf{S}'=\mathcal{R}_\varphi\mathbf{S}$
by rotating the coordinates by an angle $\varphi$, where 
\begin{equation}\label{eqn:rot}
 \mathcal{R}_\varphi\! =\!
 \begin{pmatrix}
  1 & 0 & 0 & 0\\
  0 & \cos2\varphi & \sin2\varphi & 0 \\
  0 & -\sin2\varphi & \cos2\varphi & 0 \\
  0 & 0 & 0 & 1
 \end{pmatrix}\,.
\end{equation}
Defining the position angle of the plane of linear polarization, $\theta$, by $Q\pm iU=Pe^{\pm 2i\theta}$ with $P=\sqrt{Q^2+U^2}$ and $2\theta=\arctan(U/Q)$, the rotation of coordinates shifts the position angle as $\theta\to \theta'=\theta-\varphi$.

The action of any polarization-altering device on $\mathbf{S}$ can be encoded in a Mueller matrix $\mathcal{M}$, so that the outgoing Stokes vector reads $\mathbf{S}'=\mathcal{M}\mathbf{S}$ \cite{bass2009handbook}.
In our case of interest, $\mathbf{S}$ represents the incoming CMB radiation and $\mathcal{M}$ the Mueller matrix of a telescope that employs a rotating HWP as a polarization modulator, i.e.\
\begin{equation}\label{eqn:Mueller_model}
    \mathbf{S}' = \mathcal{M}_\text{det} \mathcal{R}_{\xi-\phi} \mathcal{M}_\textsc{hwp} \mathcal{R}_{\phi+\psi}\mathbf{S}\,,
\end{equation}
where $\mathcal{R}_\varphi$ is given in eq.~\eqref{eqn:rot}. 
The meaning of each angle appearing in eq.\ \eqref{eqn:Mueller_model} is clarified in figure \ref{fig:angles}. For example,  $\mathcal{R}_{\phi+\psi}$ rotates the sky coordinates by an angle $\psi$ to the telescope coordinates (the left panel) and further rotates by $\phi$ to the HWP coordinates  (the middle panel). Here, $\mathcal{M}_\text{det}$ and $\mathcal{M}_\textsc{hwp}$ are the Mueller matrices of a detector along $x_\text{det}$ and of a general HWP:
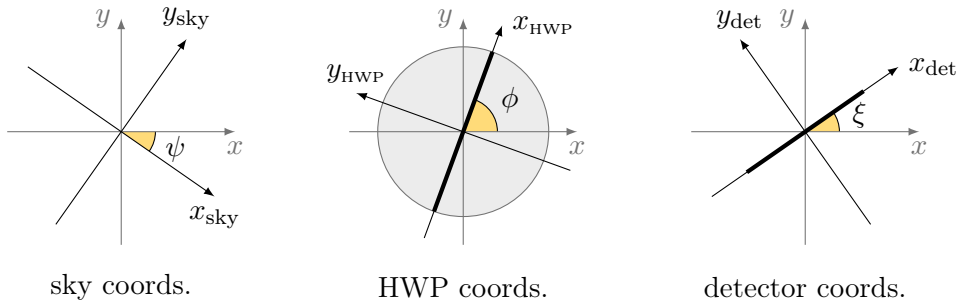
\begin{figure}[t]
    \centering
    \def\skyang{-35}
    \def\HWPang{70}
    \def\detang{35}
    \begin{tikzpicture}[scale=0.75]
    \fill[yellow!50!pink] (0,0) -- (0.6,0) arc (0:\skyang:0.6) -- (0,0);
    \draw (0.6,0) arc (0:\skyang:0.6);
    \draw[-latex, black!55] (-2,0) -- (2,0) node[below] {$x$};
    \draw[-latex, black!55] (0,-2) -- (0,2) node[left] {$y$};
    \begin{scope} [rotate=\skyang]
        \draw[-latex] (-2,0) -- (2,0) node[below] {$x_\text{sky}$};
        \draw[-latex] (0,-2) -- (0,2) node[above] {$y_\text{sky}$};
    \end{scope}
    \node at (\skyang/2:1) {$\psi$};
    \node at (0,-2.75) {sky coords.};
    \begin{scope}[shift={(6,0)}]
        \fill[gray!15] (0,0) circle (1.5);
        \fill[yellow!50!pink] (0,0) -- (0.6,0) arc (0:\HWPang:0.6) -- (0,0);
        \draw (0.6,0) arc (0:\HWPang:0.6);
        \draw[black!55] (0,0) circle (1.5);
        \draw[-latex, black!55] (-2,0) -- (2,0) node[below] {$x$};
        \draw[-latex, black!55] (0,-2) -- (0,2) node[left] {$y$};
        \begin{scope} [rotate=\HWPang]
            \draw[-latex] (-2,0) -- (2,0) node[right] {$x_\textsc{hwp}$};
            \draw[-latex] (0,-2) -- (0,2) node[above] {$y_\textsc{hwp}$};
            \draw[ultra thick] (-1.5,0) -- (1.5,0);
        \end{scope}
        \node at (\HWPang/2:1) {$\phi$};
        \node at (0,-2.75) {HWP coords.};
    \end{scope}
    \begin{scope}[shift={(12,0)}]
        \fill[yellow!50!pink] (0,0) -- (0.6,0) arc (0:\detang:0.6) -- (0,0);
        \draw (0.6,0) arc (0:\detang:0.6);
        \draw[-latex, black!55] (-2,0) -- (2,0) node[below] {$x$};
        \draw[-latex, black!55] (0,-2) -- (0,2) node[left] {$y$};
        \begin{scope} [rotate=\detang]
            \draw[-latex] (-2,0) -- (2,0) node[right] {$x_\text{det}$};
            \draw[-latex] (0,-2) -- (0,2) node[above] {$y_\text{det}$};
            \draw[ultra thick] (-1.25,0) -- (1.25,0);
        \end{scope}
        \node at (\detang/2:1) {$\xi$};
        \node at (0,-2.75) {detector coords.};
    \end{scope}
    \end{tikzpicture}
    \caption{
    The $\mathbf{S}$ vector is defined in sky coordinates, forming an angle $\psi$ with the telescope ones (left panel). The HWP optical axis and the detector's sensitive direction are rotated with respect to the telescope coordinates by angles $\phi$ and $\xi$, respectively (center and right panels). The angles are defined in right-handed coordinates with the $z$ axis taken in the direction of the telescope boresight.
    }\label{fig:angles}
\end{figure}
 \begin{equation}\label{eqn:ideal}
 \mathcal{M}_\text{det}\! =\!
 \frac{1}{2}\begin{pmatrix}
  1 & 1 & 0 & 0\\
  1 & 1 & 0 & 0\\
  0 & 0 & 0 & 0\\
  0 & 0 & 0 & 0
 \end{pmatrix}\!,\hspace{2.0225mm}\,\,
 \mathcal{M}_\textsc{hwp}\! =\!
 \begin{pmatrix}
  m_\textsc{ii} & m_\textsc{iq} & m_\textsc{iu} & m_\textsc{iv} \\
  m_\textsc{qi} & m_\textsc{qq} & m_\textsc{qu} & m_\textsc{qv} \\
  m_\textsc{ui} & m_\textsc{uq} & m_\textsc{uu} & m_\textsc{uv}\\
  m_\textsc{vi} & m_\textsc{vq} & m_\textsc{vu} & m_\textsc{vv}
 \end{pmatrix}\!.
\end{equation}
We can then model the signal $d$ measured by one detector as
\begin{equation}\label{eqn:data_model}
    d = \mathbf{a}^T\!\mathcal{M}_\text{det} \mathcal{R}_{\xi-\phi} \mathcal{M}_\textsc{hwp} \mathcal{R}_{\phi+\psi} \mathbf{S} + n\,, \quad \text{with}\quad \mathbf{a}^T=\begin{pmatrix}
     1 & 0 & 0 & 0
    \end{pmatrix},
\end{equation}
where $n$ represents an additional noise term.
\paragraph{Modeling the TOD}
In a realistic CMB experiment, $n_\text{det}$ detectors collect data by scanning the sky for an extended period of time, resulting in $n_\text{obs}$ observations for each detector. All together, these $n_\text{det}\times n_\text{obs}$ measurements constitute the TOD. We represent the TOD as a vector $\mathbf{d}$ given by
\begin{equation}\label{eqn:data_model_TOD}
    \mathbf{d} = A \mathbf{m} + \mathbf{n}\,,
\end{equation}
where $\mathbf{m}$ denotes the $\{I,Q,U,V\}$ pixelized sky maps, $A$ the response matrix, and $\mathbf{n}$ the noise component. Eq.\ \eqref{eqn:data_model_TOD} generalizes eq.\ \eqref{eqn:data_model} to larger $n_\text{obs}$ and $n_\text{det}$.
\paragraph{Bin averaging map-maker}
To extract physical information from the TOD, we convert them to the map domain via some map-making procedure. A simple method often employed in the CMB analysis is the bin averaging \cite{Tegmark:1996qs}, that estimates the sky map as
\begin{equation}\label{eqn:mapmaker}
    \widehat{\mathbf{m}}=\left(\widehat{A}^T\widehat{A}\right)^{-1} \widehat{A}^T\mathbf{d}\,,
\end{equation}
where $\widehat{A}$ is the response matrix assumed by the map-maker. 
As long as the beam is axisymmetric and purely co-polarized, and the correlated component of the noise, such as $1/f$, is negligible, the bin averaging can, in principle, recover the input $\{I,Q,U,V\}$ maps. Whether the reconstructed maps actually reproduce the sky signal or not depends on how well the instrument specifics are encoded in the map-maker or, in other words, how close $\widehat{A}$ is to $A$. When $\widehat{A}=A$ and $\mathbf{n}$ is uncorrelated in time, $\widehat{\mathbf{m}}$ is the optimal (unbiased and minimum-variance) estimator of $\mathbf{m}$.

\section{Simulation setup and output}\label{sec:simulation}
We generate statistically isotropic random Gaussian $\{I,Q,U\}$ CMB maps with \texttt{HEALPix}\footnote{\url{ http://healpix.sf.net}} \cite{Gorski:2004by} resolution of $n_\text{side}=512$ (high enough to avoid aliasing effects) by feeding the best-fit 2018 \emph{Planck} power spectra \cite{Planck:2018cosmpar} to the \texttt{synfast} function of \texttt{healpy}\footnote{\url{https://github.com/healpy/healpy}} (the Python implementation of \texttt{HEALPix}). We choose to neglect $V$ here, since the circularly polarized component of CMB is expected to be negligible\footnote{In the standard cosmological model, no circular polarization can be produced at last scattering. A number of models that could source $V$ have been proposed (see for instance \cite{Cooray:2002nm,Alexander:2008fp,Bavarsad:2009hm,Sadegh:2017rnr,Inomata:2018vbu,Vahedi:2018abn,Alexander:2019sqb,Bartolo:2019eac,Lembo:2020ufn}), but none of them predicts a strong signal, making $V\equiv 0$ a good first approximation.}.

\begin{figure}[t]
    \centering
    \includegraphics{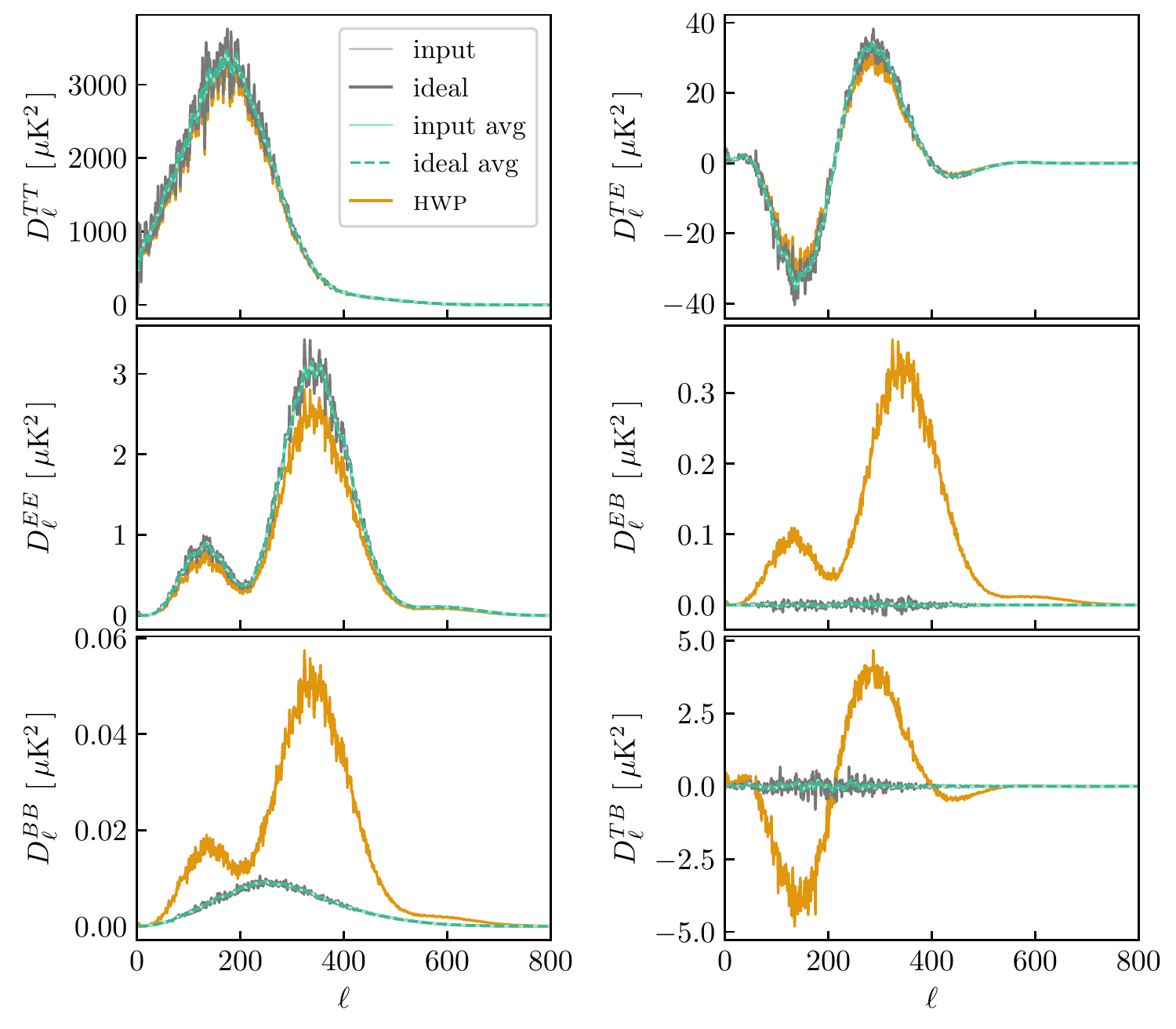}
    \caption{
    Comparison of the input angular power spectra $D_{\ell,\text{in}}^{XY}$ (light gray) with the ones computed from the outputs of the TOD simulations with ideal ($D_{\ell,\text{ideal}}^{XY}$, in dark gray) and non-ideal HWP ($D_{\ell,\textsc{hwp}}^{XY}$, in orange).
    The inputs are hard to see, since they almost perfectly overlap with the $D_{\ell,\text{ideal}}^{XY}$, while the $D_{\ell,\textsc{hwp}}^{XY}$ show clear deviations from the inputs. For clarity, we also show the simple moving average over 7 multipoles of $D_{\ell,\text{in}}^{XY}$ (lighter teal) and $D_{\ell,\text{ideal}}^{XY}$ (darker teal, dashed): $D_{\ell,\text{avg}}^{XY} \equiv \frac{1}{7}\sum_{\ell'=\ell-3}^{\ell+3} D^{XY}_{\ell'}$. The beam transfer function is not deconvolved.}
    \label{fig:inputVSidealVSnon}
\end{figure}

The observation of the input maps is simulated by a modified version of the publicly available library \texttt{beamconv}. 
This choice is motivated by \texttt{beamconv}'s ability to simulate TOD with realistic HWPs, scanning strategies and beams, which makes it a promising framework to develop simulations for, among others, LiteBIRD-like experiments. The changes we have implemented to the library all aim to better tailor the simulations to LiteBIRD-like specifics. In particular:
\begin{table}[t]
    \centering
    \begin{tabular}{p{0.29\textwidth}p{0.14\textwidth}}
        \arrayrulecolor{gray!70}
        \multicolumn{2}{l}{\vspace{-2mm}Scanning strategy parameters \vspace{2.5mm}}\\
        \hline
        \vspace{-2mm} Precession angle & \vspace{-2mm} $45^\circ$\\
        Boresight angle & $50^\circ$\\
        Precession period & $192.348$ min\\
        Spin rate & $0.05$ rpm \\
        \textcolor{white}{B}
    \end{tabular}
    \hfill
    \begin{tabular}{p{0.29\textwidth}p{0.14\textwidth}}
        \arrayrulecolor{gray!70}
       \multicolumn{2}{l}{\vspace{-2mm}Instrument properties \vspace{2.5mm}}\\
        \hline
        \vspace{-2mm} Number of detectors & \vspace{-2mm} $160$\\
        MFT frequency channel & $140$ GHz\\
        Sampling frequency & 19 Hz\\
        HWP rotation rate & 39 rpm\\
        Beam FWHM & 30.8 arcmin
    \end{tabular}
    \caption{Simulation parameters used in this work. All values are taken from \cite{LiteBIRD:2022cnt}, except for the number of detectors and the central frequency, which we choose arbitrarily.}
    \label{tab:specs}
\end{table}
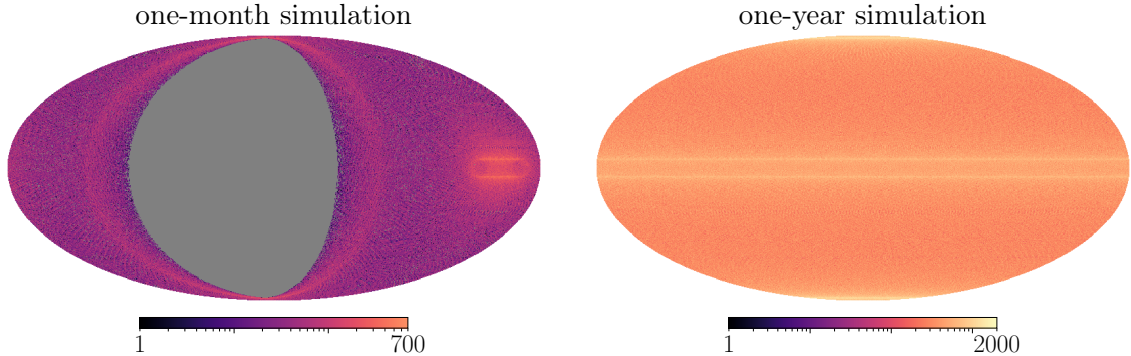
\begin{figure}[t]
    \centering
    \begin{tikzpicture}
    \node at (0,-2.3) {\includegraphics[width=.475\textwidth]{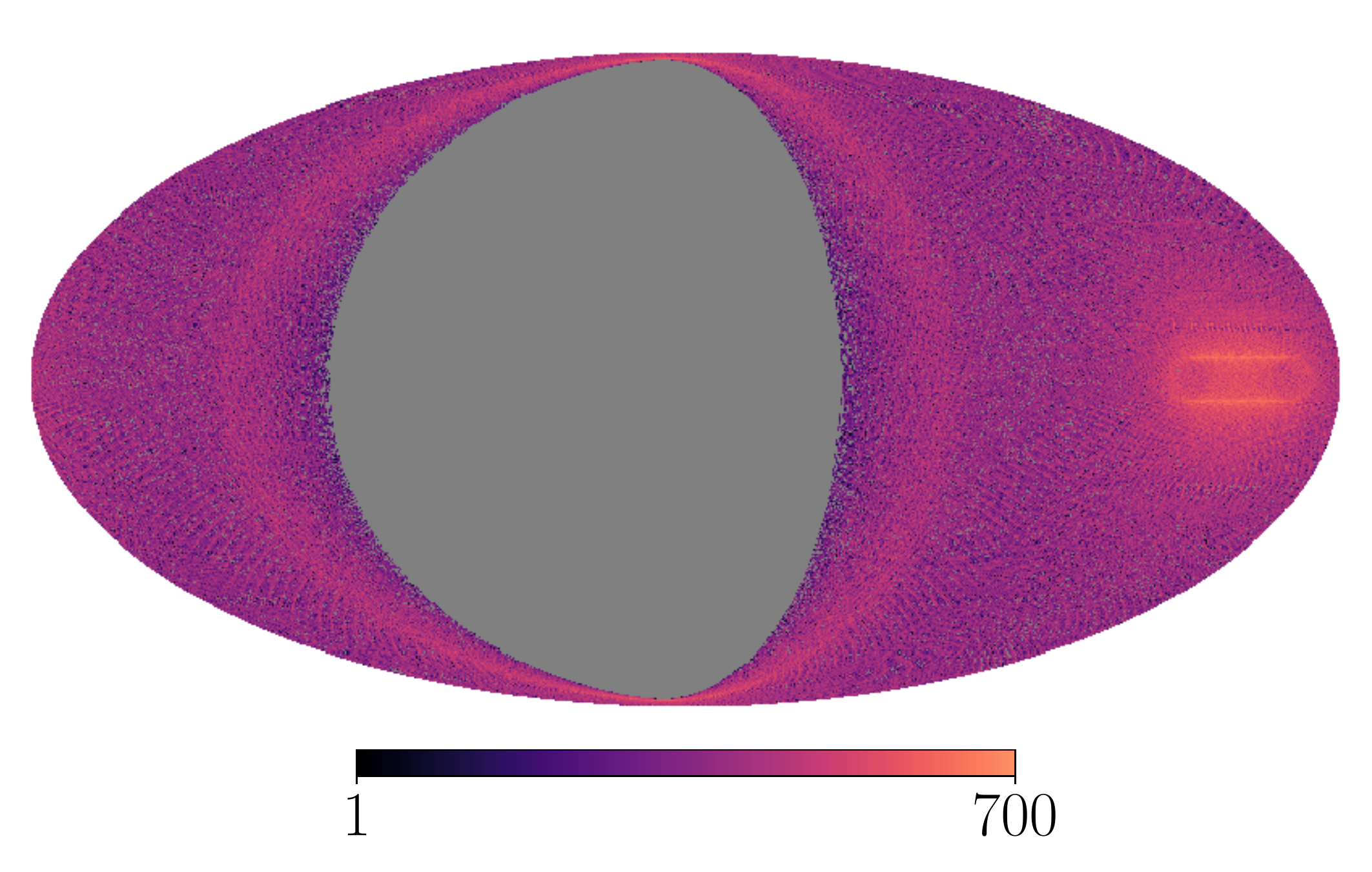}};
    \node at (0,0) {\textcolor{white}{y}one-month simulation\textcolor{white}{y}};
    \node at (7.75,-2.3) {\includegraphics[width=.475\textwidth]{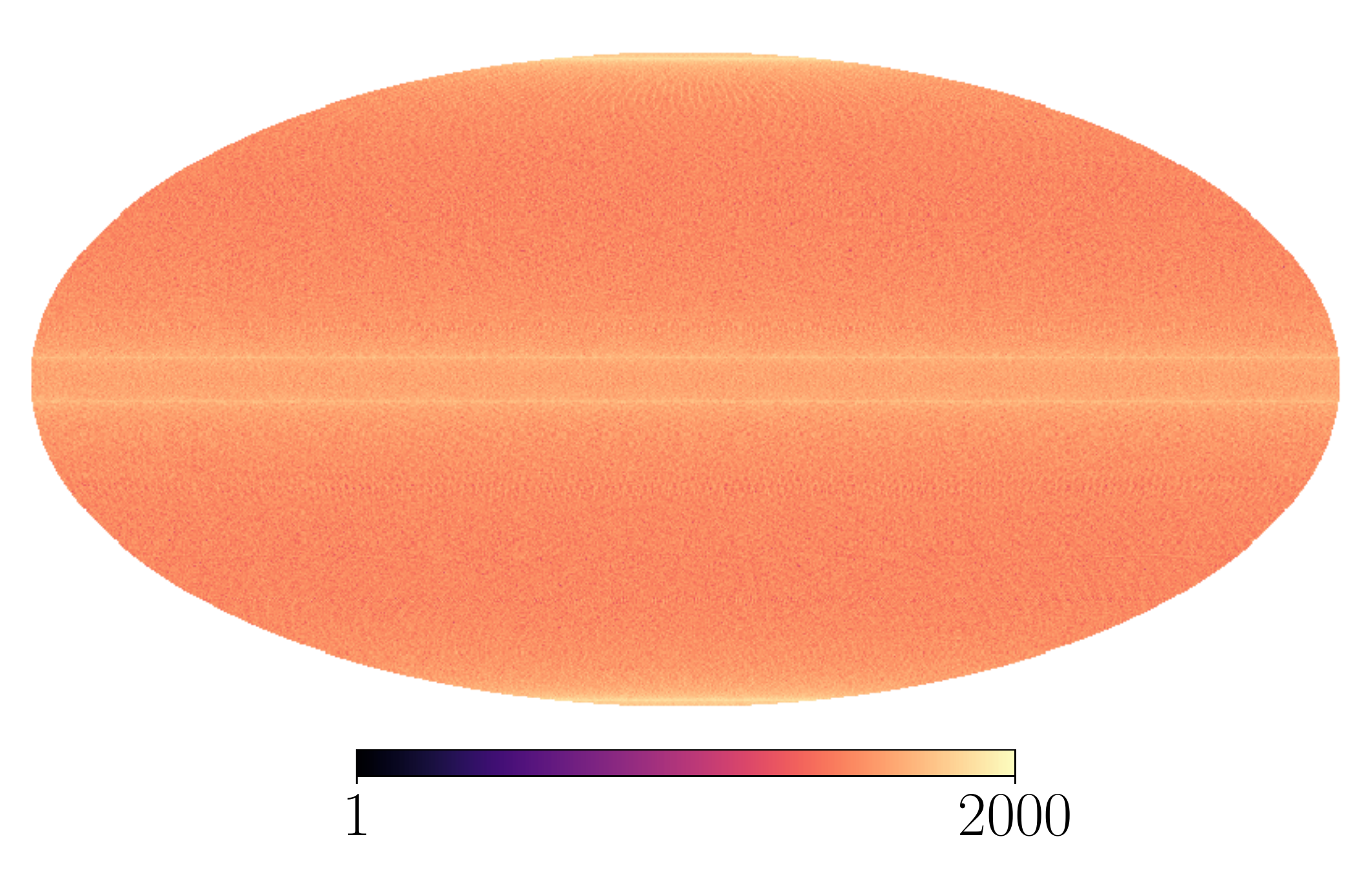}};
    \node at (7.75,0) {one-year simulation};
    \foreach \i in {2,...,9}
    {
        \draw[ultra thin] ({1.7625*(2*log10(\i)/log10(700)-1)},-4.125) -- ({1.7625*(2*log10(\i)/log10(700)-1)},-4.165);
        \draw[ultra thin] ({1.7625*(2*log10(\i*10)/log10(700)-1)},-4.125) -- ({1.7625*(2*log10(\i*10)/log10(700)-1)},-4.165);
        \ifthenelse{\i<7}{
            \draw[ultra thin] ({1.7625*(2*log10(\i*100)/log10(700)-1)},-4.125) -- ({1.7625*(2*log10(\i*100)/log10(700)-1)},-4.165);
            }{}
    }
    \foreach \i in {1,...,2}
    {
        \draw[ultra thin] ({1.7625*(2*log10(10^\i)/log10(700)-1)},-4.125) -- ({1.7625*(2*log10(10^\i)/log10(700)-1)},-4.19);
    }
    \begin{scope}[xshift=7.75cm]
    \foreach \i in {2,...,9}
    {
        \draw[ultra thin] ({1.7625*(2*log10(\i)/log10(2000)-1)},-4.125) -- ({1.7625*(2*log10(\i)/log10(2000)-1)},-4.165);
        \draw[ultra thin] ({1.7625*(2*log10(\i*10)/log10(2000)-1)},-4.125) -- ({1.7625*(2*log10(\i*10)/log10(2000)-1)},-4.165);
        \draw[ultra thin] ({1.7625*(2*log10(\i*100)/log10(2000)-1)},-4.125) -- ({1.7625*(2*log10(\i*100)/log10(2000)-1)},-4.165);
    }
    \foreach \i in {1,...,3}
    {
        \draw[ultra thin] ({1.7625*(2*log10(10^\i)/log10(2000)-1)},-4.125) -- ({1.7625*(2*log10(10^\i)/log10(2000)-1)},-4.19);
    }
    \end{scope}
    \end{tikzpicture}
    \caption{Simulated boresight hit maps for one-month (left) and one-year (right) observations. The two panels share the same logarithmic color map, although the range shown is truncated for the one-month case. Unobserved pixels appear gray.}
    \label{fig:scanning}
\end{figure}
\begin{description}[topsep=2mm,parsep=1mm,itemsep=1mm]
\item[Scanning strategy]
We implement a LiteBIRD-like scanning strategy by mimicking the relevant functionalities of \texttt{pyScan}\footnote{\url{https://github.com/tmatsumu/LB_SYSPL_v4.2}}. The values of the telescope boresight and precession angles, together with their rotation parameters, are specified in table \ref{tab:specs}. We simulate one year of observations to cover the full sky (see figure \ref{fig:scanning}).
\item[Instrument]
We work with 160 detectors from the 140 GHz channel of LiteBIRD's Medium Frequency Telescope (MFT) and read the relevant parameters from \cite{LiteBIRD:2022cnt}: the HWP rotation rate, the full-width-at-half-maximum (FWHM) of the (Gaussian and co-polarized) beam, the instrument sampling frequency and the detectors’ pointing offsets. See table \ref{tab:specs} for their numerical values.
\item[HWP Mueller matrix] The Mueller matrix elements for the MFT's HWP at 140 GHz are taken from \cite{Giardiello:2021uxq}, up to a coordinate change from International Astronomical Union (IAU) to CMB standards that flips the sign of the $m_\textsc{iu}$, $m_\textsc{qu}$, $m_\textsc{ui}$ and $m_\textsc{uq}$ elements:
\begin{equation}\label{eqn:muellerHWP}
    \mathcal{M}_\textsc{hwp} = 
    \begin{pmatrix*}[l]
    \phantom{-}9.80\times10^{-1} & \,\phantom{-}\!1.81\times10^{-2}   & \,-9.81\times10^{-3} \\
    \phantom{-}1.81\times10^{-2}   & \,\phantom{-}\!9.71\times10^{-1} & \,-1.21\times10^{-1} \\
    -9.81\times10^{-3}          & \,-1.21\times10^{-1}          & \,-8.40\times10^{-1} 
    \end{pmatrix*}\,.
\end{equation}
This is the HWP Mueller matrix we assume when including non-idealities\footnote{Doing so, we neglect the dependence of the HWP properties on the angle of incidence. The consequences of such approximation have not been tested yet.}. Since the elements of $\mathcal{M}_\textsc{hwp}$ are frequency-dependent, choosing a different frequency would result in slightly different output spectra.
\end{description}
We run two simulations for one-year observations. 
Noise is not included in either simulation to isolate the effect of HWP non-idealities in the signal; thus, using a different $n_\text{det}$ is almost free from consequences and our results do not change using fewer detectors. In the first simulation we assume the ideal HWP by setting $\mathcal{M}_\textsc{hwp}=\mathcal{M}_\text{ideal}\equiv\text{diag}(1,1,-1)$, while we account for non-idealities in the second one. We convert both TODs to $\{{I},{Q},{U}\}$ maps by the bin averaging map-maker (see eq.\ \eqref{eqn:mapmaker}) whose response matrix $\widehat{A}$ assumes the ideal HWP described by $\mathcal{M}_\text{ideal}$. We then calculate two sets of full-sky angular power spectra using the \texttt{anafast} function of \texttt{healpy}. We denote the first (second) set of output spectra with $C_{\ell,\text{ideal}}^{XY}$ ($C_{\ell,\textsc{hwp}}^{XY}$), where $X,Y=\{T,E,B\}$.
The rescaled $D_{\ell,\text{ideal}}^{XY} \equiv \ell(\ell+1) C_{\ell,\text{ideal}}^{XY}/2\pi$ and $D_{\ell,\textsc{hwp}}^{XY}$ spectra are plotted in figure \ref{fig:inputVSidealVSnon}, together with the input spectra multiplied by the Gaussian beam transfer functions, $D_{\ell,\text{in}}^{XY}$. 
The simple map-maker recovers the input spectra with average deviations less than 0.1\% in the plotted range when processing the TOD generated with $\mathcal{M}_\text{ideal}$, while important discrepancies arise for the non-ideal case.

We do not account for photometric calibration, although it represents a crucial step in any CMB analysis pipeline. Gain calibration, if  perfect, would ensure intensity to be recovered exactly, hence compensating the lack of power in $D_\ell^{TT}$ visible in figure \ref{fig:inputVSidealVSnon}. The discrepancies in $D_\ell^{TE}$ and $D_\ell^{TB}$ would also be reduced, although not removed. The discussion and results presented in the following would however not change, reason why we omit the step. 
\section{Analytical estimate of the output spectra}\label{sec:analytical}
To understand the simulation results, we derive approximate analytical formulae for the angular power spectra affected by HWP non-idealities.
Since we are neglecting any circularly polarized component, the Stokes vector is given by ${\mathbf{S}}=({I},{Q},{U})$. To obtain analytical formulae we apply the bin averaging map-maker of eq.\ \eqref{eqn:mapmaker} to a minimal TOD consisting of the signals measured by four detectors with different polarization sensitivity directions\footnote{This is the minimal configuration that can reconstruct linearly polarized radiation.} (with $0^\circ$, $90^\circ$, $45^\circ$ and $135^\circ$ offsets) observing the same sky pixel. By expressing the signals observed by each of the four detectors as functions of the input Stokes parameters according to eq.\ \eqref{eqn:data_model}, we obtain
\begin{subequations}\label{eqn:IQUhat}
\begin{align}
    \widehat{I} &= 
    m_\textsc{ii} I_\text{in}+ (m_\textsc{iq} Q_\text{in}+m_\textsc{iu} U_\text{in})\cos (2 \alpha) + (m_\textsc{iq} U_\text{in}-m_\textsc{iu} Q_\text{in})\sin (2 \alpha) \,,\\
    \widehat{Q} &=
    \frac{1}{2} \Bigl\{(m_\textsc{qq}-m_\textsc{uu})Q_\text{in}+ (m_\textsc{qu}+m_\textsc{uq})U_\text{in} + 2m_\textsc{qi} I_\text{in}\cos (2 \alpha) + 2m_\textsc{ui} I_\text{in}\sin (2 \alpha)\nonumber\\
    &\qquad\quad+\bigl[(m_\textsc{qq}+m_\textsc{uu})Q_\text{in} + (m_\textsc{qu}-m_\textsc{uq})U_\text{in}\bigr]\cos (4 \alpha) \nonumber\\
    &\qquad\quad+ \bigl[-(m_\textsc{qu}-m_\textsc{uq})Q_\text{in}+(m_\textsc{qq}+m_\textsc{uu})U_\text{in}\bigr] \sin (4 \alpha)\Bigr\}\,,\label{eqn:Qhat}\\
    \widehat{U} &=
    \frac{1}{2} \Bigl\{ (m_\textsc{qq}-m_\textsc{uu})U_\text{in}- (m_\textsc{qu}+m_\textsc{uq})Q_\text{in} - 2m_\textsc{ui} I_\text{in}\cos (2 \alpha) + 2m_\textsc{qi} I_\text{in}\sin (2 \alpha)\nonumber\\
    &\qquad\quad+ \bigl[- (m_\textsc{qq}+m_\textsc{uu})U_\text{in}+(m_\textsc{qu}-m_\textsc{uq})Q_\text{in}\bigr]\cos (4 \alpha) \nonumber\\
    &\qquad\quad+ \bigl[ (m_\textsc{qu}-m_\textsc{uq})U_\text{in}+(m_\textsc{qq}+m_\textsc{uu})Q_\text{in}\bigr]\sin (4 \alpha)\Bigr\}\,,\label{eqn:Uhat}
\end{align}
\end{subequations}
where $m_{\textsc{s}\textsc{s}'}$ $(\textsc{s},\textsc{s}'=\textsc{i,q,u})$ are the elements of non-ideal $\mathcal{M}_\textsc{hwp}$ and
$\alpha$ denotes the sum of the HWP's ($\phi$) and the telescope's ($\psi$) angles\footnote{\label{foot:expl}For the simple 4-detector configuration we are considering, the response matrix can be expressed as $A=B\mathcal{R}_{\xi-\phi} \mathcal{M}_\textsc{hwp} \mathcal{R}_{\phi+\psi}$, where $B$ accounts for the different $\xi$ angles of the four detectors and happens to satisfy $B^TB = \text{diag}(1,1/2,1/2)$. As for the map-maker response matrix, $\widehat{A}=B\mathcal{R}_{\xi-\phi} \mathcal{M}_\text{ideal} \mathcal{R}_{\phi+\psi}$. All $B\mathcal{R}_{\xi-\phi}$ terms cancel out in eq.\ \eqref{eqn:mapmaker} and we are left with $\widehat{\mathbf{S}}=\mathcal{R}_{\phi+\psi}^T \mathcal{M}_\text{ideal}\mathcal{M}_\textsc{hwp}\mathcal{R}_{\phi+\psi}\mathbf{S}_\text{in}$. The discrepancies between $\widehat{\mathbf{S}}$ and $\mathbf{S}_\text{in}$ can therefore only depend on $\phi+\psi$.}: $\alpha\equiv \phi+\psi$. The quantities with the subscript ``in'' on the right hand side denote the sky signals, whereas ${\widehat{\mathbf{S}}}=(\widehat{I},\widehat{Q},\widehat{U})$ on the left hand side are maps recovered by the map-maker. These formulae are applicable to our case as long as eq.\ \eqref{eqn:data_model} accurately models the TOD simulated by \texttt{beamconv}, which is the case for an axisymmetric and purely co-polarized beam.  

Eqs.\ \eqref{eqn:IQUhat} model $\widehat{\mathbf{S}}_p$ reconstructed from four observations of the pixel $p$, one for each detector. If each of those 4 detectors were to observe that same pixel $n_p$ times, the change in eqs.\ \eqref{eqn:IQUhat} would amount to substituting
\begin{equation}\label{eqn:morenp}
  \cos (n\alpha) \to \frac{1}{n_p} \sum_{t=t_1}^{t_{n_p}} \cos (n \alpha_t)\,,\quad \sin (n\alpha) \to \frac{1}{n_p} \sum_{t=t_1}^{t_{n_p}} \sin (n \alpha_t)\,,
\end{equation}
for $n=\{2,4\}$. If $p$ is observed with a uniform enough sample of $\alpha_t$ values and $n_p$ is large enough, these terms can be neglected, resulting in
\begin{equation}\label{eqn:reconstructed}
    \widehat{\mathbf{S}} \simeq \begin{pmatrix}
    m_{\textsc{ii}} I_\text{in} \\
    [(m_{\textsc{qq}}-m_{\textsc{uu}})Q_\text{in} + (m_{\textsc{qu}}+m_{\textsc{uq}})U_\text{in}]/2 \\
    [(m_{\textsc{qq}}-m_{\textsc{uu}})U_\text{in}-(m_{\textsc{qu}}+m_{\textsc{uq}})Q_\text{in}]/2
    \end{pmatrix}\,.
\end{equation}
We expect this to be a good approximation, given the presence of a rapidly spinning HWP and the good coverage of the simulation (see figure \ref{fig:scanning}).

\begin{figure}[t]
    \centering
    \includegraphics{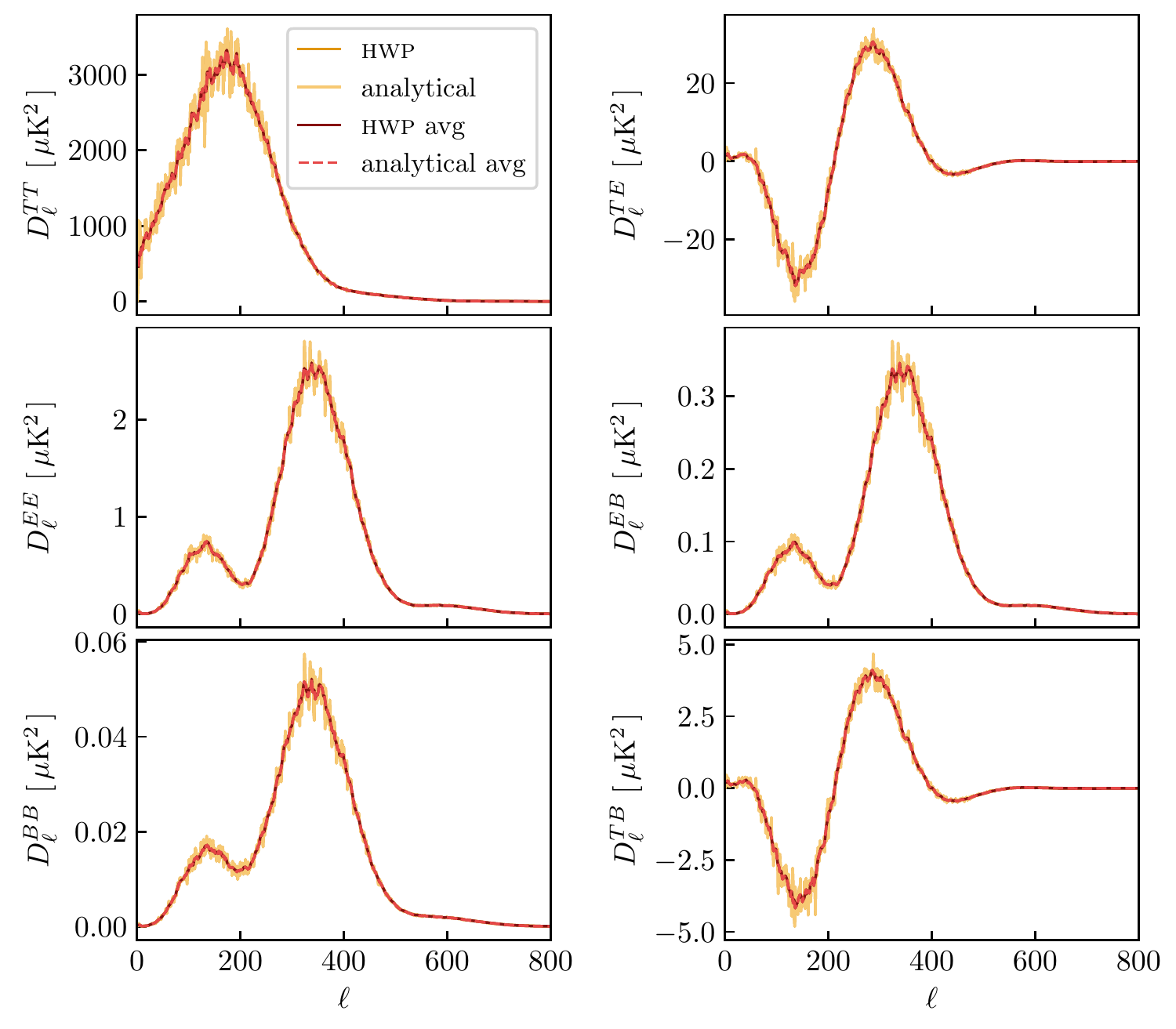}
    \caption{
    Comparison of the spectra computed from the output of the TOD simulation with non-ideal HWP, $D_{\ell,\textsc{hwp}}^{XY}$ (dark orange), with the $\widehat{D}_\ell^{XY}$ from the analytical formulae given in eqs.\ \eqref{eqn:theospectra} (light orange). The non-ideal outputs are hard to see, since they almost perfectly overlap with the analytical curves. For clarity, we also show the simple moving average over 7 multipoles of $D_{\ell,\textsc{hwp}}^{XY}$ (dark red) and $\widehat{D}_\ell^{XY}$ (light red, dashed): $D_{\ell,\text{avg}}^{XY} \equiv \frac{1}{7}\sum_{\ell'=\ell-3}^{\ell+3} D^{XY}_{\ell'}$. The beam transfer function is not deconvolved.}
    \label{fig:inputVStheoVSnon}
\end{figure}

By expanding eq.\ \eqref{eqn:reconstructed} in spherical harmonics, we write the corresponding angular power spectra as a mixing of the input ones weighted by combinations of the non-ideal HWP's Mueller matrix elements:
\begin{subequations}\label{eqn:theospectra}
\begin{align}
    \widehat{C}_{\ell}^{TT}\!&\simeq m_\textsc{ii}^2 C_{\ell,\text{in}}^{TT},\\
    \widehat{C}_{\ell}^{EE}\!&\simeq \frac{(m_\textsc{qq}-m_\textsc{uu})^2}{4} C_{\ell,\text{in}}^{EE} + \frac{(m_\textsc{qu}+m_\textsc{uq})^2}{4} C_{\ell,\text{in}}^{BB} + \frac{(m_\textsc{qq}-m_\textsc{uu})(m_\textsc{qu}+m_\textsc{uq})}{2} C_{\ell,\text{in}}^{EB},\\
    \widehat{C}_{\ell}^{BB}\!&\simeq  \frac{(m_\textsc{qq}-m_\textsc{uu})^2}{4} C_{\ell,\text{in}}^{BB} + \frac{(m_\textsc{qu}+m_\textsc{uq})^2}{4} C_{\ell,\text{in}}^{EE} - \frac{(m_\textsc{qq}-m_\textsc{uu})(m_\textsc{qu}+m_\textsc{uq})}{2} C_{\ell,\text{in}}^{EB},\\
    \widehat{C}_{\ell}^{TE}\!&\simeq \frac{m_\textsc{ii}(m_\textsc{qq}-m_\textsc{uu})}{2}C_{\ell,\text{in}}^{TE} +  \frac{m_\textsc{ii}(m_\textsc{qu}+m_\textsc{uq})}{2}C_{\ell,\text{in}}^{TB},\\
    \widehat{C}_{\ell}^{EB}\!&\simeq \frac{(m_\textsc{qq}\!-\hspace{-.075em}m_\textsc{uu})^2\!-\hspace{-.075em}(m_\textsc{qu}\!+\hspace{-.075em}m_\textsc{uq})^2}{4} C_{\ell,\text{in}}^{EB} \!-\hspace{-.075em} \frac{(m_\textsc{qq}\!-\hspace{-.075em}m_\textsc{uu})(m_\textsc{qu}\!+\hspace{-.075em}m_\textsc{uq})}{4} (C_{\ell,\text{in}}^{EE}\!-\hspace{-.075em}C_{\ell,\text{in}}^{BB}),\\
    \widehat{C}_{\ell}^{TB}\!&\simeq  \frac{m_\textsc{ii}(m_\textsc{qq}-m_\textsc{uu})}{2}C_{\ell,\text{in}}^{TB} -  \frac{m_\textsc{ii}(m_\textsc{qu}+m_\textsc{uq})}{2}C_{\ell,\text{in}}^{TE}.
\end{align}
\end{subequations}
These analytical formulae explain quite well the non-ideal output spectra $C_{\ell,\textsc{hwp}}^{XY}$ (see figure \ref{fig:inputVStheoVSnon}). They are especially accurate on large scales, $\ell\lesssim 500$, where average deviations between $C_{\ell,\textsc{hwp}}^{XY}$ and $\widehat{C}_\ell$ are less than $0.1\%$. Larger deviations on smaller scales are due to the approximate nature of eq.\ \eqref{eqn:reconstructed}. Cosine and sine terms do not average out exactly, resulting in pixel-by-pixel fluctuations on smaller scales.
\section{Impact on cosmic birefringence}\label{sec:impact}
Next generation CMB experiments are expected to measure the CMB polarization with unprecedented sensitivity and improve the constraints on the CB angle, $\beta$, recently obtained from the \emph{Planck} data \cite{Minami:2020odp,Diego-Palazuelos:2022dsq,Eskilt:2022wav,Eskilt:2022cff}. Here we discuss how HWP non-idealities can impact such constraints in the particular case of a LiteBIRD-like mission discussed so far.

\begin{figure}[t]
    \centering
    \includegraphics{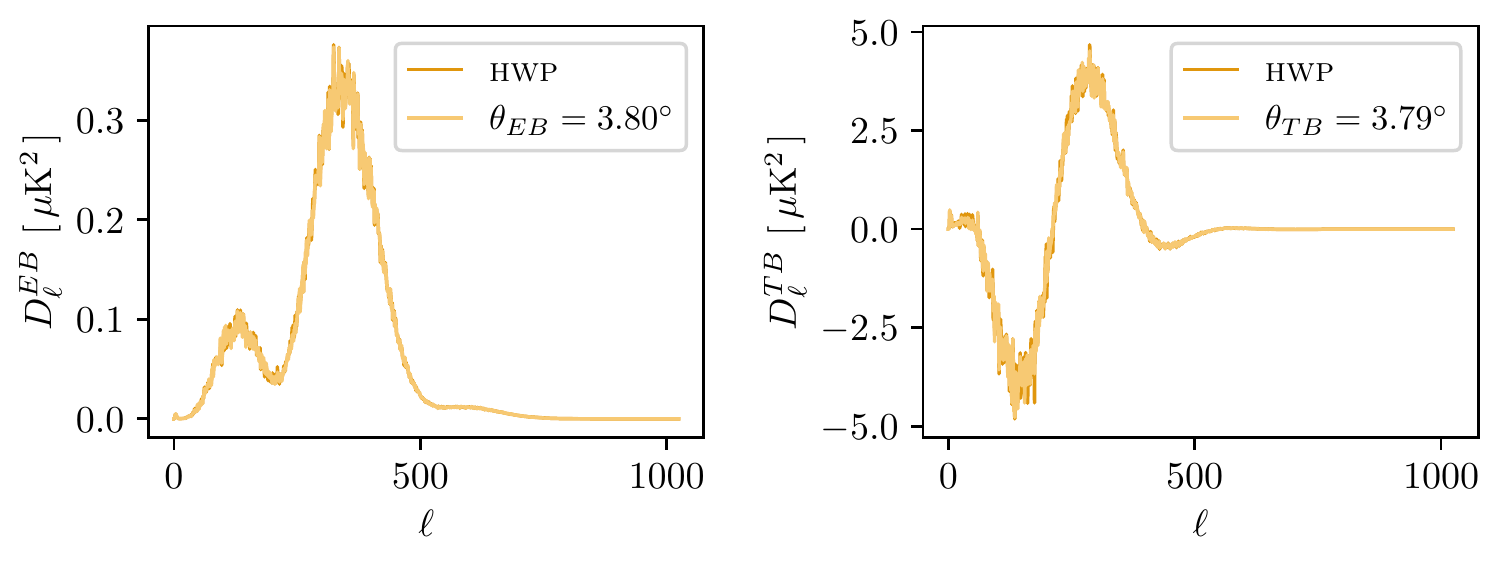}
    \caption{
    Comparison of $D_{\ell,\textsc{hwp}}^{EB}$ and $D_{\ell,\textsc{hwp}}^{TB}$ computed from the outputs of the TOD simulation with non-ideal HWP (dark orange) with the best-fit estimates of $\tan (4\theta_{EB}) (D_{\ell,\textsc{hwp}}^{EE}-D_{\ell,\textsc{hwp}}^{BB})/2$ and $\tan(2\theta_{TB})D_{\ell,\textsc{hwp}}^{TE}$, respectively.}
    \label{fig:fits}
\end{figure}

First, we recall that the sign of $\beta$ reported in the literature is also chosen to follow the CMB convention and a positive $\beta$ corresponds to a clockwise rotation on the sky \cite{Komatsu:2022nvu}. The isotropic CB angle, $\beta$, and a miscalibration of the instrument polarization angle, $\Delta\alpha$, affect the observed spectra identically, since both rotate the observed Stokes parameters in the same way. 
The observed spectra then satisfy the equations \cite{lue,feng}
\begin{equation}\label{eqn:degenerate}
    C_{\ell,\text{obs}}^{EB} = \frac{\tan(4\theta)}{2}\left(C_{\ell,\text{obs}}^{EE}-C_{\ell,\text{obs}}^{BB}\right)\,, \qquad
    C_{\ell,\text{obs}}^{TB} = \tan(2\theta)C_{\ell,\text{obs}}^{TE}\,,
\end{equation}
where $\theta$ represents rotation in the position angle of the plane of linear polarization including $\beta$, $\Delta \alpha$, or their sum. 
Not accounting for the HWP non-idealities in the map-maker step is degenerate with $\theta$, as it is evident from both our simulations and the analytical formulae given in eq.\ \eqref{eqn:theospectra}. We will refer to this additional rotation of the polarization plane as the HWP-induced miscalibration.
\paragraph{HWP-induced miscalibration from the simulated output spectra}
We separately fit the simulated $C_{\ell,\textsc{hwp}}^{EB}$ and $C_{\ell,\textsc{hwp}}^{TB}$ for the angles $\theta_{EB}$ and $\theta_{TB}$, respectively, using the least-squares method with variance given by 
\begin{equation}\label{eqn:cosmicvar}
    \text{Var}\bigl(C_{\ell,\textsc{hwp}}^{XY}\bigr) = \frac{1}{2\ell+1}\left[C_{\ell,\textsc{hwp}}^{XX}C_{\ell,\textsc{hwp}}^{YY}+\bigl(C_{\ell,\textsc{hwp}}^{XY}\bigr)^2\right]\,,
\end{equation}
for $XY=\{EB,TB\}$, respectively. The best-fit values, $\theta_{EB}=3.800^\circ\pm 0.007^\circ$ and $\theta_{TB}=3.79^\circ\pm 0.02^\circ$, are compatible with each other in agreement with eqs.\ \eqref{eqn:degenerate}. The observed and best-fit spectra are plotted in figure \ref{fig:fits} and are in good agreement.
\paragraph{HWP-induced miscalibration from the analytical formulae}
Using the fact that both $C_{\ell_\text{in}}^{EB}$ and $C_{\ell_\text{in}}^{TB}$ simply fluctuate around zero, eqs.\ \eqref{eqn:theospectra} can be rearranged to express $\widehat{C}_\ell^{EB}$ and $\widehat{C}_\ell^{TB}$ similarly to the $C_{\ell,\text{obs}}^{XY}$ of eqs.\ \eqref{eqn:degenerate}:
\begin{equation}\label{eqn:exp_degenerate}
    \widehat{C}_{\ell}^{EB} \simeq \frac{\tan(4\widehat\theta)}{2}\left(\widehat{C}_{\ell}^{EE}-\widehat{C}_{\ell}^{BB}\right)\,, \qquad
    \widehat{C}_{\ell}^{TB} \simeq \tan(2\widehat\theta)\widehat{C}_{\ell}^{TE}\,,
\end{equation}
where
\begin{equation}\label{eqn:widehattheta}
    \widehat{\theta} = -\frac{1}{2}\arctan\left(\frac{m_\textsc{qu}+m_\textsc{uq}}{m_\textsc{qq}-m_\textsc{uu}}\right)\simeq 3.8^\circ\,,
\end{equation}
in agreement with the best-fit values reported above.

If we were to relax all the underlying assumptions at once, we could not write $\widehat{\theta}$ this compactly. However, controlled generalizations do not necessarily spoil the simplicity of the analytical formulae. For instance, accounting for the frequency dependence of both the HWP Mueller matrix elements and the CMB signal, $\widehat{\theta}$ can be expressed as (see appendix \ref{app:freq} for the derivation):
\begin{equation}\label{eqn:widehattheta_freq}
    \widehat{\theta} = -\frac{1}{2}\arctan\left(\frac{\int \text{d}\nu\, S_\text{CMB}(\nu)\left[ m_\textsc{qu}+m_\textsc{uq}\right](\nu)}{\int \text{d}\nu\, S_\text{CMB}(\nu)\left[ m_\textsc{qq}-m_\textsc{uu}\right](\nu)}\right),
\end{equation}
where $S_\text{CMB}(\nu)$ denotes the CMB spectral energy distribution (SED). 

Another assumption that can be relaxed without spoiling the simplicity of the analytical formulae is the absence of miscalibration angles in the map-maker. When the telescope, HWP, and detector angles are not exactly known, $\psi=\widehat{\psi}+\delta\psi$, $\phi=\widehat{\phi}+\delta\phi$, and $\xi=\widehat{\xi}+\delta\xi$, where the hat denotes the values assumed by the map-maker.  As long as we neglect the frequency dependence of $\delta\psi$, $\delta\phi$ and $\delta\xi$, we find (see appendix \ref{app:misc} for the derivation)
\begin{equation}\label{eqn:widehattheta_misc}
    \widehat{\theta} = -\frac{1}{2}\arctan\left(\frac{\int \text{d}\nu\, S_\text{CMB}(\nu)\left[ m_\textsc{qu}+m_\textsc{uq}\right](\nu)}{\int \text{d}\nu\, S_\text{CMB}(\nu)\left[ m_\textsc{qq}-m_\textsc{uu}\right](\nu)}\right)+\delta \xi -\delta \psi -2 \delta \phi\,.
\end{equation}
The sign difference between the contributions from $\delta\xi$ and $\delta\psi+2\delta\phi$ is due to the presence of the HWP. Ideally, the HWP acts on a polarization vector by reflecting it over its fast axis. This causes counterclockwise rotations applied before the HWP to look clockwise after, meaning that $\delta\phi+\delta\psi$ should be subtracted from  $\delta\xi-\delta\phi$ (see eq.\ \eqref{eqn:Mueller_model}).
\section{Conclusions and outlook}\label{sec:conclusions}
In this work, we studied how overlooking HWP non-idealities during map-making can affect the reconstructed angular power spectra of CMB temperature and polarization fields. We focused on the impact of non-idealities on the measurement of the CB angle, $\beta$.

As a concrete working case, we considered a single frequency channel (140 GHz) of a space CMB mission with LiteBIRD-like specifics: scanning strategy, sampling frequency, detectors' pointing offsets and their polarization sensitivity directions, FWHM of the Gaussian beam and HWP specifics (rotation frequency and Mueller matrix elements). We employed the publicly available beam-convolution code \texttt{beamconv} to simulate the noiseless TOD for the above instrument and scanning specifications. We ran two different simulations: the HWP has been assumed to be ideal in the first simulation, while a realistic Mueller matrix has been employed in the second. We then converted both TODs to maps by a bin averaging map-maker that neglects the HWP non-idealities.
As expected, the output spectra computed from the ideal simulation faithfully recovered the input spectra, while the spectra of the non-ideal maps showed a very different behavior (figure \ref{fig:inputVSidealVSnon}). We also derived a set of analytical formulae (see eq.\ \eqref{eqn:theospectra}) that accurately model the reconstructed angular power spectra as functions of the input spectra and the HWP Mueller matrix elements.

We studied the impact of the HWP non-idealities on $\beta$. We found that neglecting them in the map-making step induces an additional miscalibration of the polarization angle which might be erroneously interpreted as CB. For the concrete case we studied, the miscalibration angle induced by the HWP non-idealities amounts to $\theta\simeq3.8^\circ$. This value, obtained by fitting the output angular power spectra from the simulation, is compatible with the prediction from the analytical formulae (see eq.\ \eqref{eqn:widehattheta}).

Definitive confirmation of the current hint of CB \cite{Minami:2020odp,Diego-Palazuelos:2022dsq,Eskilt:2022wav,Eskilt:2022cff} requires the systematic uncertainty in the absolute position angle of linear polarization to be well below $0.1^\circ$ \cite{Komatsu:2022nvu}. We must therefore acquire accurate knowledge of the Mueller matrix elements via calibration, so that the systematic uncertainty in $\theta$ due to HWP non-idealities is well below $0.1^\circ$. With such knowledge, one can take into account HWP non-idealities either during the map-making step or when interpreting the angular power spectra. As one cannot know the Mueller matrix elements perfectly, any remaining mismatch between the true Mueller matrix and the matrix assumed by the map-maker still affects the power spectra. Our simulation and analytical formulae will be useful for deriving the required accuracy of HWP calibration to meet specific science goals.

The situation we considered in this paper is still simplistic: we simulated a single frequency channel in the absence of noise, and we used a Gaussian beam and a simple bin averaging map-maker. However, a similar analysis can be carried out for more complex cases. It is of utmost importance to make better predictions about how HWP non-idealities realistically affect the data collected by CMB experiments and, therefore, the cosmological information extracted from them. In this direction, we plan to carry on the following steps: i) drop the single frequency approximation, generalizing the results discussed here to a finite frequency bandwidth; ii) add a noise component to the TOD; iii)
study the combined effect of beam asymmetries and HWP non-idealities; iv) include non-idealities in the map-maker and study how the uncertainties in our knowledge of non-idealities might propagate to the observed angular power spectra; and v) derive requirements for the accuracy of HWP calibration. We leave these topics for future work.

\acknowledgments
We thank T. Matsumura for providing us with the \texttt{pyScan} code and useful discussions, and S. Giardiello for constructive comments to the manuscript.
This work was supported in part by JSPS KAKENHI grants no.\ JP20H05850 and no.\ JP20H05859, the Excellence Cluster ORIGINS which is funded by the Deutsche Forschungsgemeinschaft (DFG, German Research Foundation) under Germany’s Excellence Strategy: Grant No.~EXC-2094 - 390783311, and 
Swedish National Space Agency and Vetenskapsrådet (2019-03959). This work has also received funding from the European Union's
Horizon 2020 research and innovation programme under the Marie Skłodowska-Curie grant agreement no.\ 101007633. The Kavli IPMU is supported by World Premier International Research Center Initiative (WPI), MEXT, Japan. The Flatiron
Institute is supported by the Simons Foundation. Co-funded by the European Union (ERC, CMBeam, 101040169). Views and opinions expressed are however those of the author(s) only and do not necessarily reflect those of the European Union or the European Research Council. Neither the European Union nor the granting authority can be held responsible for them.


\appendix

\section{Finite frequency bandwidth}\label{app:freq}
Taking into account the frequency dependence of both the HWP Mueller matrix elements and the CMB signal, we write the data model of eq.\ \eqref{eqn:data_model} as
\begin{equation}\label{eqn:data_model_freq}
    d = \mathbf{a}^T\!\mathcal{M}_\text{det} \mathcal{R}_{\xi-\phi} \int\text{d}\nu\, \mathcal{M}_\textsc{hwp}(\nu) \mathcal{R}_{\phi+\psi} \mathbf{S}(\nu) + n\,.
\end{equation}
Repeating the analysis presented in section \ref{sec:analytical}, eq.\ \eqref{eqn:theospectra} reads
\begin{subequations}\label{eqn:theospectra_freq}
\begin{align}
    \widehat{C}_{\ell}^{TT}\!\!&\simeq \langle m_\textsc{ii} \rangle^2 \bar{C}_{\ell,\text{in}}^{TT},\\
    \widehat{C}_{\ell}^{EE}\!\!&\simeq \frac{\langle m_\textsc{qq}-m_\textsc{uu}\rangle ^2}{4} \bar{C}_{\ell,\text{in}}^{EE} + \frac{\langle m_\textsc{qu}+m_\textsc{uq}\rangle ^2}{4} \bar{C}_{\ell,\text{in}}^{BB} + \frac{\langle m_\textsc{qq}-m_\textsc{uu}\rangle \langle m_\textsc{qu}+m_\textsc{uq}\rangle }{2} \bar{C}_{\ell,\text{in}}^{EB},\!\\
    \widehat{C}_{\ell}^{BB}\!\!&\simeq  \frac{\langle m_\textsc{qq}-m_\textsc{uu}\rangle ^2}{4} \bar{C}_{\ell,\text{in}}^{BB} + \frac{\langle m_\textsc{qu}+m_\textsc{uq}\rangle ^2}{4} \bar{C}_{\ell,\text{in}}^{EE} - \frac{\langle m_\textsc{qq}-m_\textsc{uu}\rangle \langle m_\textsc{qu}+m_\textsc{uq}\rangle }{2} \bar{C}_{\ell,\text{in}}^{EB},\\
    \widehat{C}_{\ell}^{TE}\!\!&\simeq \frac{\langle m_\textsc{ii}\rangle \langle m_\textsc{qq}-m_\textsc{uu}\rangle }{2}\bar{C}_{\ell,\text{in}}^{TE} +  \frac{\langle m_\textsc{ii}\rangle \langle m_\textsc{qu}+m_\textsc{uq}\rangle }{2}\bar{C}_{\ell,\text{in}}^{TB},\\
    \widehat{C}_{\ell}^{EB}\!\!&\simeq \frac{\langle m_\textsc{qq}\!-\hspace{-.075em}m_\textsc{uu}\rangle ^2\!-\hspace{-.075em}\langle m_\textsc{qu}\!+\hspace{-.075em}m_\textsc{uq}\rangle ^2}{4} \bar{C}_{\ell,\text{in}}^{EB} \!-\hspace{-.075em} \frac{\langle m_\textsc{qq}\!-\hspace{-.075em}m_\textsc{uu}\rangle \langle m_\textsc{qu}\!+\hspace{-.075em}m_\textsc{uq}\rangle }{4} ( \bar{C}_{\ell,\text{in}}^{EE}\!-\hspace{-.075em}\bar{C}_{\ell,\text{in}}^{BB}),\\
    \widehat{C}_{\ell}^{TB}\!\!&\simeq  \frac{\langle m_\textsc{ii}\rangle \langle m_\textsc{qq}-m_\textsc{uu}\rangle }{2}\bar{C}_{\ell,\text{in}}^{TB} -  \frac{\langle m_\textsc{ii}\rangle \langle m_\textsc{qu}+m_\textsc{uq}\rangle }{2}\bar{C}_{\ell,\text{in}}^{TE},
\end{align}
\end{subequations}
where the brackets denote frequency integrals weighted over the SED of the CMB,
\begin{equation}
    \langle f\rangle\equiv \frac{\int \text{d}\nu\, S_\text{CMB}(\nu) f(\nu)}{\int \text{d}\nu\, S_\text{CMB}(\nu) }\,,
\end{equation} 
and $\bar{C}_{\ell,\text{in}}^{XY}$ the input angular power spectra at some reference frequency $\bar{\nu}$. This modifies eq.\ \eqref{eqn:widehattheta} to
\begin{equation}\label{eqn:widehattheta_freq_APP}
    \widehat{\theta} = -\frac{1}{2}\arctan\left(\frac{\int \text{d}\nu\, S_\text{CMB}(\nu)\left[ m_\textsc{qu}+m_\textsc{uq}\right](\nu)}{\int \text{d}\nu\, S_\text{CMB}(\nu)\left[ m_\textsc{qq}-m_\textsc{uu}\right](\nu)}\right).
\end{equation}

\section{Additional miscalibration angles}\label{app:misc}
So far, we neglected any miscalibration angles in the map-maker, i.e.\ we assumed the response matrix $\widehat{A}$ to encode the true values of the telescope, HWP, and detector angles: $\widehat{\psi}\equiv \psi$, $\widehat{\phi}\equiv \phi$, and $\widehat{\xi}\equiv \xi$, where the hat denotes the values assumed by the map-maker. We now consider a more general case by allowing for deviations: $\psi=\widehat{\psi}+\delta\psi$, $\phi=\widehat{\phi}+\delta\phi$, and $\xi=\widehat{\xi}+\delta\xi$.
\paragraph{Single frequency}
Repeating the analysis presented in section \ref{sec:analytical} with miscalibration angles, eq.\ \eqref{eqn:reconstructed} reads
\begin{subequations}
\begin{align}
    \widehat{I} &\simeq m_{\textsc{ii}} I_\text{in}\,, \\
    \widehat{Q} &\simeq [\cos(2\delta\theta)(m_{\textsc{qq}}-m_{\textsc{uu}})+\sin(2\delta\theta)(m_{\textsc{qu}}+m_{\textsc{uq}})]Q_\text{in}/2 \nonumber \\
    &\qquad + [\cos(2\delta\theta)(m_{\textsc{qu}}+m_{\textsc{uq}})-\sin(2\delta\theta)(m_{\textsc{qq}}-m_{\textsc{uu}})]U_\text{in}/2 \,, \\
    \widehat{U} &\simeq
    [\cos(2\delta\theta)(m_{\textsc{qq}}-m_{\textsc{uu}})+\sin(2\delta\theta)(m_{\textsc{qu}}+m_{\textsc{uq}})]U_\text{in}/2 \nonumber \\
    &\qquad - [\cos(2\delta\theta)(m_{\textsc{qu}}+m_{\textsc{uq}})+\sin(2\delta\theta)(m_{\textsc{qq}}-m_{\textsc{uu}})]Q_\text{in}/2\,,
\end{align}
\end{subequations}
where $\delta\theta\equiv \delta \xi -\delta \psi -2 \delta \phi$. This modifies eq.\ \eqref{eqn:widehattheta} to
\begin{eqnarray}
\nonumber
    \widehat{\theta} &=& -\frac{1}{2}\arctan\left(\frac{\cos(2\delta\theta)(m_{\textsc{qu}}+m_{\textsc{uq}})-\sin(2\delta\theta)(m_{\textsc{qq}}-m_{\textsc{uu}})}{\cos(2\delta\theta)(m_{\textsc{qq}}-m_{\textsc{uu}})+\sin(2\delta\theta)(m_{\textsc{qu}}+m_{\textsc{uq}})}\right)\\
    &=&-\frac{1}{2}\arctan\left(\frac{m_\textsc{qu}+m_\textsc{uq}}{m_\textsc{qq}-m_\textsc{uu}}\right)+\delta\theta\,.
\end{eqnarray}
Therefore, the additional miscalibration angles simply shift $\widehat{\theta}$, as expected.

\paragraph{Finite frequency bandwidth}
Taking into account a finite frequency bandwidth and miscalibration angles simultaneously is slightly more complicated, but does not spoil the analytic treatment as long as $\delta\theta$ is assumed to be frequency-independent. The generalization of eq.\ \eqref{eqn:widehattheta} in this case reads
\begin{equation}
    \widehat{\theta} = -\frac{1}{2}\arctan\left(\frac{\int \text{d}\nu\, S_\text{CMB}(\nu)\left[ m_\textsc{qu}+m_\textsc{uq}\right](\nu)}{\int \text{d}\nu\, S_\text{CMB}(\nu)\left[ m_\textsc{qq}-m_\textsc{uu}\right](\nu)}\right)+\delta\theta\,.
\end{equation}

\bibliographystyle{JHEP}
\bibliography{bibliography}

\end{document}